\documentclass[showpacs,prl,twocolumn,superscriptaddress]{revtex4}
\usepackage{epsfig}% Include figure files
\usepackage{bm}% bold math

\begin{document}
\title{Mott Insulators in the Strong Spin-Orbit Coupling Limit: \\
From Heisenberg to a Quantum Compass and Kitaev Models}
\author{G.~Jackeli}
\altaffiliation[]{Also at E.~Andronikashvili Institute of Physics, 0177 
Tbilisi, Georgia
}
\affiliation{Max-Planck-Institut f\"ur Festk\"orperforschung, 
Heisenbergstrasse 1, D-70569 Stuttgart, Germany}
\author{G.~Khaliullin}
\affiliation{Max-Planck-Institut f\"ur Festk\"orperforschung, 
Heisenbergstrasse 1, D-70569 Stuttgart, Germany}
\begin{abstract}
We study the magnetic interactions in Mott-Hubbard 
systems with partially filled $t_{2g}$-levels and with strong spin-orbit
coupling. The latter entangles the spin and orbital spaces, and leads to a 
rich variety of the low energy Hamiltonians that extrapolate from the 
Heisenberg to a quantum compass model depending on the lattice geometry.
This gives way to ``engineer'' in such Mott insulators an exactly solvable 
spin model by Kitaev relevant for quantum computation.  
We, finally, explain ``weak'' ferromagnetism, with an anomalously large 
ferromagnetic moment, in Sr$_2$IrO$_4$.
\end{abstract}
\date{\today}
\pacs{
75.30.Et, %Exchange and superexchange interactions
71.70.Ej,  %Spin-orbit coupling, Zeeman and Stark splitting, Jahn-Teller effect
75.10.Jm %Quantized spin models
} 
\maketitle 

The transition metal compounds with partially filled $d$-levels have been the
subject of extensive studies after the discovery  
of a variety of novel physical phenomena and a diversity of new  phases 
\cite{Ima98,Tok00,Dag05}. In the undoped compounds a strong Coulomb 
repulsion localizes the $d$-electrons in Mott-Hubbard or charge-transfer
insulating  regimes \cite{Zaa85}. The low energy physics of 
such insulators, in some cases, are described in terms of spin-only
Hamiltonians. This happens when the symmetry of the local surrounding of
a transition metal (TM) ion is low enough to lift the orbital degeneracy of 
a $d$-level as in case of, {\it e.g.}, high-$T_c$ cuprates. However, often, 
a TM ion possesses an orbital degeneracy in addition to that 
originating from spin. Typically, the orbitals form a 
long-range-ordered pattern driven by Jahn-Teller or exchange interactions, 
and, being subject to a discrete symmetry, behave more like static
(classical) objects compared to their spin partners. 
Magnetic properties of orbitally ordered systems depend 
on the symmetry of the occupied orbitals. Indeed, orbital ordering may 
stabilize various types of magnetic phases \cite{Kug82}, 
as well as spin gapped states without any long-range spin order 
\cite{Pen97,Jac08}. In other circumstances, the situation can be  opposite: 
the orbitals may remain in a liquid state down to the lowest temperatures
(due to strong  quantum fluctuations), while the spins are 
slowly fluctuating about a long range ordered state 
\cite{Kei00,Kha00}.

In this Letter we want to discuss yet another situation, when a strong 
relativistic spin-orbit (SO) coupling entangles locally the 
spin and orbital degrees of freedom. The physics of such systems  
may drastically differ from that of compounds in which spin-orbit 
coupling is of a perturbative nature, as the form of magnetic 
interactions is no longer dictated by a global spin $SU(2)$ symmetry alone. 
The effects of a strong SO interaction on cooperative magnetic phenomena 
has been discussed in the pioneering works by Kanamori on 
Fe$^{2+}$ and Co$^{2+}$ compounds \cite{Kan56}. In recent years, 
there has been a revival of interest in spin-orbit coupling in 
the context of exchange interactions \cite{Kha04,Kha05,Che08}, magnetoelectric 
\cite{Kat05} and spin Hall effects \cite {Mur04}, Fermi-surface topology 
\cite{Hav08}, {\it etc}.  

The SO coupling is strongly enhanced for the late transition 
metal ions such as Ir, Os, Rh, Ru. Indeed, optical absorption data on 
Ir$^{4+}$ impurities in SrTiO$_3$ suggests a fairly high value of the SO 
coupling $\lambda \sim 380$ meV \cite{Sch84}. This far exceeds possible 
intersite interactions between the $t_{2g}$ orbitals and spins 
in the insulating iridates, and hence is able to lock them together 
forming a total angular momentum locally. 
In the following we focus on the systems composed of 
magnetic ions with a single hole in a threefold degenerate
$t_{2g}$-level, a low spin state of $d^5$-configuration, such as 
Ir$^{4+}$ or Rh$^{4+}$ ions subject to a strong octahedral field.
We formulate a superexchange theory for such systems and show that 
together with conventional interactions of predominantly Heisenberg form, 
more exotic spin models such as the quantum compass model naturally appear 
as low energy Hamiltonians. We suggest how to implement in such Mott 
insulators an exactly solvable model proposed by Kitaev \cite{Kit06}, 
which exhibit exotic anyonic excitations with fractional statistics.
We apply the present theory to the insulating iridium compound Sr$_2$IrO$_4$
\cite{Hua94,Cao98,Moo06,Kim08} exhibiting ``weak'' ferromagnetism (FM) 
with an anomalously large FM moment. 

\begin{figure}
\epsfysize=20mm
\centerline{\epsffile{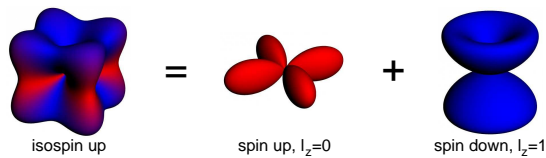}}
\caption{(Color online) Density profile of a hole in the isospin up state 
(without tetragonal distortion). It is a superposition of a spin up hole 
density in $|xy\rangle$-orbital, $l_z=0$, (middle) and spin down one in 
$(|yz\rangle+i|xz\rangle)$ state, $l_z=1$, (right).}
\label{fig1}
\end{figure}

{\it Single ion Kramers doublet}.-- We first introduce the local magnetic 
degrees of freedom. In the low spin $d^5$ configuration a hole resides 
in $t_{2g}$ manifold of $xy$, $xz$, $yz$ orbitals, and has an effective 
angular momentum $l=1$ \cite{Abr70}:
$|l_z\!\!=\!\!0\rangle \equiv \!\!|xy\rangle$, 
$|l_z\!\!=\!\!\pm 1\rangle 
\equiv \!\!-\frac{1}{\sqrt{2}}(i|xz\rangle\pm|yz\rangle)$. 
The total moment ${\vec M}=2{\vec s}-{\vec l}$, where ${\vec s}$ is a 
hole spin operator. The single ion Hamiltonian 
$H_{0}=\lambda {\vec l}\cdot{\vec s}+\Delta l_z^2$ consists of a SO 
coupling with $\lambda>0$ and a possible tetragonal splitting $\Delta$ of 
the $t_{2g}$ levels. $\Delta>0$ for an oxygen octahedron elongated 
along the $z\parallel c$-axis. The lowest energy level of $H_{0}$ 
is a Kramers doublet of isospin states 
$|\tilde{\uparrow}\rangle$ and $|\tilde{\downarrow}\rangle$: 
\begin{eqnarray}
|\tilde{\uparrow}\rangle&=&\sin\theta |0,\uparrow\rangle
-\cos\theta |+1,\downarrow\rangle~,
\nonumber\\
|\tilde{\downarrow}\rangle&=&\sin\theta |0,\downarrow\rangle
-\cos\theta |-1,\uparrow\rangle~.
\label{eq1}
\end{eqnarray} 
Angle $\theta$ parameterizes the relative strength of the tetragonal 
splitting, with $\tan(2\theta)=2\sqrt{2}\lambda/(\lambda-2\Delta)$. 
Notice that the wave functions of the Kramers doublet 
are given by a coherent superposition of different orbital 
and spin states, leading to a peculiar distribution of spin densities in real 
space (see Fig.~\ref{fig1}). This will have important consequences 
for the symmetry of the intersite interactions. Namely, the very 
form of the exchange Hamiltonian depends on bond geometry through a density 
profile of Kramers states, as we demonstrate below.  
%%%%%%%%%%%%%%%%%%%%%%%%%%%%%%%%%%%%%%%%%%%%%%%%%%%%%%%%%%%%%%%%%%%%%%%%%%%%
%%%%%%%%%%%%%%%%%%%%%%%%%%%%%%%%%%%%%%%%%%%%%%%%%%%%%%%%%%%%%%%%%%%%%%%%%%%%

\begin{figure}
\epsfysize=50mm
\centerline{\epsffile{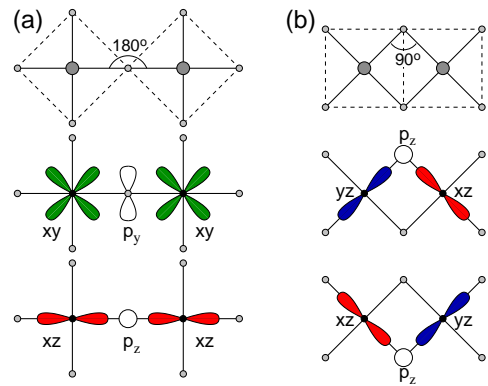}}
\caption{(Color online) Two possible geometries of a TM-O-TM bond with 
corresponding orbitals active along these bonds. The large (small) dots 
stand for the transition metal (oxygen) ions. (a) 180$^\circ$-bond formed by 
corner shared octahedra, and (b) 90$^\circ$-bond formed by edge 
shared octahedra.}
\label{fig2}
\end{figure}

{\it Exchange couplings of neighboring Kramers states}.-- 
We consider the limit of the strong 
spin-orbit coupling, {\it i.e.}, when $\lambda$ is larger than exchange 
interaction between the isospins. 
The exchange Hamiltonians for isospins are then obtained 
by projecting the corresponding superexchange spin-orbital models onto the 
isospin states Eq.~(\ref{eq1}). First, we present the results for the case 
of cubic symmetry ($\Delta=0,~\sin\theta=1/\sqrt{3}$), and discuss later 
the effects of a tetragonal distortion. We consider two 
common cases of TM-O-TM bond geometries: (A) 180$^\circ$-bond formed 
by corner-shared octahedra as in Fig.~\ref{fig2}(a), and (B) 90$^\circ$-bond 
formed by edge-shared ones, Fig.~\ref{fig2}(b).

(A) 180$^\circ$-bond: For this geometry, the nearest-neighbor $t_{2g}$ 
hopping matrix is diagonal in the orbital space and, on a given bond, 
only two orbitals are active, {\it e.g.}, $|xy\rangle$ and $|xz\rangle$ 
orbitals along a bond in $x$-direction [Fig.~\ref{fig1}(a)]. The spin-orbital 
exchange Hamiltonian for such a system has already been reported: see  
Eq.~(3.11) in Ref.~\cite{Kha05}. After projecting it onto the ground state 
doublet, we find an exchange Hamiltonian for isospins in a form of Heisenberg 
plus a pseudo-dipolar interaction: 
\begin{equation}
{\cal H}_{ij}=J_1\vec S_i\cdot \vec S_{j}+J_2(\vec S_{i}
\cdot\vec r_{ij})(\vec r_{ij}\cdot\vec S_{j})~,
\label{eq2}
\end{equation}
where $\vec S_i$ is the $S=1/2$ operator for isospins 
(referred to as simply spins from now on), $\vec r_{ij}$ is the unit 
vector along the $ij$ bond, and $J_{1(2)}=\frac{4}{9}\nu_{1(2)}$. Hereafter,
we use the energy scale $4t^2/U$ where $t$ is a $dd$-transfer integral 
through an intermediate oxygen, and $U$ stands for the Coulomb repulsion 
on the same orbitals. The parameters $\nu_{1(2)}$ controlling isotropic
(anisotropic) couplings are given by $\nu_1=(3r_1+r_2+2r_3)/6$ and 
$\nu_2=(r_1-r_2)/4$, where the set of $r_n$ characterizing the multiplet 
structure of the excited states depends solely on 
the ratio $\eta=J_{H}/U$ of Hund's coupling and $U$ \cite{note1}. 
At small $\eta$, one has $\nu_1\simeq 1$ and $\nu_2\simeq \eta/2$. 
Thus, we find a predominantly isotropic Hamiltonian,
with a weak dipolar-like anisotropy term. While the overall form of 
Eq. (\ref{eq2}) could be anticipated from symmetry arguments, 
the explicit derivation led us to an unexpected result: 
In the limit of strong SO coupling, the magnetic degrees are governed by 
a nearly Heisenberg model just like in the case of small $\lambda$, 
and, surprisingly enough, its anisotropy is entirely due to the Hund's 
coupling. This is opposite to a conventional situation: typically, 
the anisotropy corrections are obtained in powers of $\lambda$ while 
the Hund's coupling is not essential. 

%%%%%%%%%%%%%%%%%%%%%%%%%%%%%%%%%%%%%%%%%%%%%%%%%%%%%%%%%%%%%%%%%%%%%%%%%%%%
%%%%%%%%%%%%%%%%%%%%%%%%%%%%%%%%%%%%%%%%%%%%%%%%%%%%%%%%%%%%%%%%%%%%%%%%%%%%%%%

(B) 90$^\circ$-bond: There are again only two orbitals active on a given 
bond, {\it e.g.}, $|xz\rangle$ and $|yz\rangle$ orbitals along 
a bond in the $xy$-plane. However, the hopping matrix has now only 
non-diagonal elements and there are two possible paths for a charge transfer 
[via upper or lower oxygen, see Fig.~\ref{fig2}(b)]. This peculiarity of a 
90$^\circ$-bond leads to an exchange Hamiltonian 
drastically different from that of a 180$^\circ$ geometry. 
Two transfer amplitudes via upper and lower oxygen interfere in a destructive 
manner and the isotropic part of the Hamiltonian {\it exactly} 
vanishes. The finite, {\it anisotropic} interaction appears, however,
due to the $J_H$-multiplet structure of the excited levels. Most importantly, 
the very form of the exchange interaction depends on the spatial orientation of
a given bond. We label a bond $ij$ laying in the $\alpha\beta$ plane 
perpendicular to the $\gamma (=x,y,z)$ axis by a $(\gamma)$-bond.
With this in mind, the Hamiltonian can be written as: 
\begin{equation}
{\cal H}_{ij}^{(\gamma)}=-JS_{i}^{\gamma}S_{j}^{\gamma}~,
\label{eq3}
\end{equation}
with $J=\frac{4}{3}\nu_2$. Remarkably, this Hamiltonian is precisely a 
quantum analog  of the so-called {\it compass model}. The latter, introduced 
originally for the orbital degrees of freedom in Jahn-Teller systems 
\cite{Kug82}, has been the subject of numerous studies as a prototype model 
with protected ground state degeneracy of topological origin 
(see, {\it e.g.}, Ref.~\onlinecite{Doc05}). However, to our knowledge,
no magnetic realization of the compass model has been proposed so far. 

{\it Implementing the Kitaev model in Mott insulators}.-- 
The Kitaev model is equivalent to a quantum compass model 
on a honeycomb lattice \cite{note2}. It shows a number of fascinating 
properties such as anyonic excitations with exotic fractional 
statistics, topological degeneracy, and, in particular, it is relevant for
quantum computation \cite{Kit06}. This generated an enormous interest in 
a possible realization of this model in real systems, with current 
proposals based on optical lattices \cite{Dua03}.
Here we outline how to ``engineer'' the Kitaev model in Mott insulators.

\begin{figure}
\epsfysize=30mm
\centerline{\epsffile{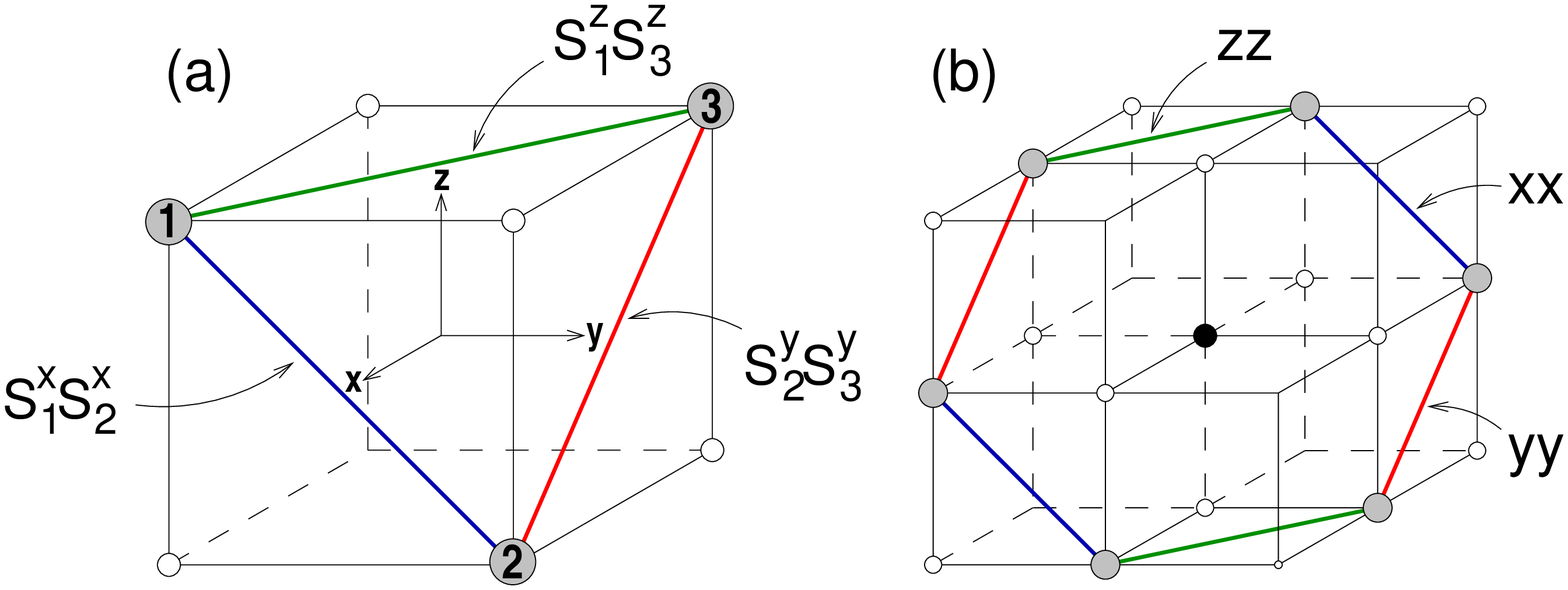}}
\caption{(Color online) Examples of the structural units formed by
90$^\circ$ TM-O-TM bonds and corresponding spin-coupling patterns.
Grey circles stand for magnetic ions, and small open circles denote 
oxygen sites. (a) Triangular unit cell of ABO$_2$-type layered compounds,
periodic sequence of this unit forms a triangular lattice of
magnetic ions. The model (\ref{eq3}) on this structure is a realization of 
a quantum compass model on a triangular lattice: {\it e.g.}, 
on a bond 1-2, laying perpendicular to $x$-axis, the interaction 
is $S_{1}^{x}S_{2}^{x}$. (b) Hexagonal unit cell of A$_2$BO$_3$-type 
layered compound, in which magnetic ions (B-sites) form a honeycomb lattice. 
(Black dot: nonmagnetic A-site). On an $xx$-bond the interaction is 
$S_{i}^{x}S_{j}^{x}$, {\it etc}. 
For this structure the model (\ref{eq3}) is identical to the Kitaev model.
}
\label{fig3}
\end{figure}

Shown in Fig.~\ref{fig3}(a) is a triangular unit formed by 90$^\circ$ bonds
together with ``compass'' interactions that follow from Eq.~(\ref{eq3}). 
Such a structure is common for a number of oxides, {\it e.g.}, 
layered compounds ABO$_2$ (where A and B are alkali  and TM ions, 
respectively). The triangular lattice of magnetic ions in an ABO$_2$ 
structure can be depleted down to a honeycomb lattice 
(by periodic replacements of TM ions with non-magnetic ones). 
One then obtains an A$_2$BO$_3$ compound, 
which has a hexagonal unit shown in Fig.~\ref{fig3}(b). 
There are three nonequivalent bonds, each being perpendicular 
to one of the cubic axes $x,y,z$. 
Then, according to Eq.~(\ref{eq3}) the spin coupling, {\it e.g.}, 
on a $(x)$-bond is of $S_i^xS_j^x$ type, precisely as in the Kitaev 
model. The honeycomb lattice provides a particularly striking example 
of new physics introduced by strong SO coupling: the Heisenberg model 
is converted into the Kitaev model with a spin-liquid ground state.  
%%%%%%%%%%%%%%%%%%%%%%%%%%%%%%%%%%%%%%%%%%%%%%%%%%%%%%%%%%%%%%%%%%%%%%%%%%%
%%%%%%%%%%%%%%%%%%%%%%%%%%%%%%%%%%%%%%%%%%%%%%%%%%%%%%%%%%%%%%%%%%%%%%%%%%%

The  compound Li$_2$RuO$_3$ \cite{Miu07} represents a physical example
of the A$_2$BO$_3$ structure. By replacement of spin-one Ru$^{4+}$ with 
spin one-half Ir$^{4+}$ ions, one may realize a strongly spin-orbit 
coupled Mott insulator with low energy physics described by the Kitaev model. 

%%%%%%%%%%%%%%%%%%%%%%%%%%%%%%%%%%%%%%%%%%%%%%%%%%%%%%%%%%%%%%%%%%%%%%%%%%%%
\begin{figure}
\epsfysize=50mm
\centerline{\epsffile{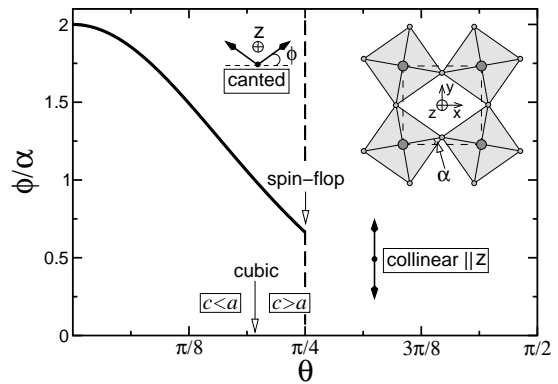}}
\caption{The spin canting angle $\phi$ (in units of $\alpha$) 
as a function of the tetragonal distortion parameter $\theta$. 
Inset shows a sketch of an IrO$_2$-plane. The oxygen octahedra 
are rotated by an angle $\pm\alpha$ about $z$-axis forming a two 
sublattice structure. In the cubic case, $\theta\simeq \pi/5$, one 
has $\phi=\alpha$ exactly. The spin-flop transition from the in-plane 
canted spin state to a collinear N\'eel ordering along 
$z$-axis occurs at $\theta=\pi/4$. 
}
\label{fig4}
\end{figure}
%%%%%%%%%%%%%%%%%%%%%%%%%%%%%%%%%%%%%%%%%%%%%%%%%%%%%%%%%%%%%%%%%%%%%%%%%%%%

{\it ``Weak'' ferromagnetism of Sr$_2$IrO$_4$}.-- 
As an example of a spin-orbit coupled Mott insulator, we discuss 
the layered compound Sr$_2$IrO$_4$, a $t_{2g}$ analog of the undoped
high-$T_c$ cuprate La$_2$CuO$_4$. In Sr$_2$IrO$_4$, a square lattice of 
Ir$^{4+}$ ions is formed by corner-sharing IrO$_6$ octahedra, 
elongated along the $c$-axis and rotated about it 
by $\alpha\simeq 11^\circ$ \cite{Hua94} (see Fig.~\ref{fig4}). 
The compound undergoes a magnetic transition at $\sim 240$~K  displaying 
a weak FM, which can be ascribed to a Dzyaloshinsky-Moriya (DM) interaction.  
The puzzling fact is that ``weak'' FM  moment is in fact unusually large,
$M_{\rm FM}\simeq 0.14 \mu_{B}$ \cite{Cao98} which is
two-orders of magnitude larger than that in La$_2$CuO$_4$ \cite{Thi88}. 
A simple estimate gives a spin canting angle $\phi\simeq 8^\circ$ which 
is close to $\alpha$, {\it i.e.}, the ordered spins seem to rigidly follow 
the staggered rotations of octahedra. Here we show that the strong SO 
coupling scenario gives a natural explanation of this observation.

We first show the dominant part of the Hamiltonian for Sr$_2$IrO$_4$ 
neglecting the Hund's coupling for a moment. Accounting for the rotations 
of IrO$_6$ octahedra, we find:  
%%%%%%%%%%%%%%%%%%%%%%%%%%%%%%%%%%%%%%%%%%%%%%%%%%%%%%%%%%%%%%%%%%%%%%%%%%%%
\begin{eqnarray}
{\cal H}=J\vec S_i\cdot \vec S_{j}+J_{z}S_{i}^{z}S_{j}^{z}
+\vec D\cdot{\big [}\vec S_i\times\vec S_j{\big ]}~.
\label{eq4}
\end{eqnarray} 
%%%%%%%%%%%%%%%%%%%%%%%%%%%%%%%%%%%%%%%%%%%%%%%%%%%%%%%%%%%%%%%%%%%%%%%%
Here, the isotropic coupling $J=\nu_1(t_s^2-t_a^2)$, where 
$t_s=\sin^2\theta+\frac{1}{2}\cos^2\theta\cos2\alpha$, 
and $t_a=\frac{1}{2}\cos^2\theta\sin2\alpha$. The second and third 
terms describe the symmetric and DM anisotropies, 
with $J_{z}=2\nu_1 t_a^2$, $\vec D=(0,0,-D)$, and $D=2\nu_1t_st_a$.
[For $\alpha=0$, these terms vanish and we recover $J_1$-term of the 
180$^\circ$ result (\ref{eq2})]. As it follows from Eq.~(\ref{eq4}), 
the spin canting angle is given by a ratio $D/J\simeq 2t_a/t_s\sim 2\alpha$ 
which is independent of $\lambda$, and is solely determined by lattice 
distortions. This explains the large spin canting angle $\phi\sim\alpha$ 
in Sr$_2$IrO$_4$. 

As in the case of weak SO coupling \cite{She92}, the Hamiltonian 
(\ref{eq4}) can in fact be mapped to the Heisenberg model $\vec
{\tilde S}_{i}\cdot\vec{\tilde S}_{j}$ where operators 
$\vec{\tilde S}$ are obtained by a staggered rotation of 
$\vec S$ around the $z$-axis by an angle $\pm\phi$, with $\tan(2\phi)=D/J$. 
Thus, at $J_H=0$, there is no true magnetic anisotropy. Once $J_H$-corrections 
are included, the Hamiltonian receives also the anisotropic terms:
\begin{equation}
{\tilde {\cal H}} =
{\tilde J}\vec {\tilde S}_i\cdot\vec{\tilde S}_{j}
-\Gamma_1{\tilde S}_{i}^{z}{\tilde S}_{j}^{z}
\pm\Gamma_2{\big (} {\tilde S}_{i}^{x}{\tilde S}_{j}^{x}
-{\tilde S}_{i}^{y}{\tilde S}_{j}^{y}{\big )},
\label{eq5}
\end{equation}
where ${\tilde J}=\nu_1(t_s^2+t_a^2)+\Gamma_1$, 
$\Gamma_1=\nu_2\cos^2\!\theta\cos2\theta$,
$\Gamma_2=\nu_2\sin^2\!\theta\cos^2\!\theta $, and a $\pm$ sign is taken for a bond
along $x(y)$-axis. This Hamiltonian supports two types of spin orderings 
(see Fig.~\ref{fig4}). For $\Gamma_1>0$ the spins form a canted structure
in $xy$-plane. We find the out-of-plane magnon gap 
$\propto \sqrt{\tilde{J}\Gamma_1}$ of a classical origin, and 
much smaller in-plane gap $\propto \Gamma_2$ generated by quantum 
fluctuations. Shown in Fig.~\ref{fig4} is the spin canting angle $\phi$ 
(in units of $\alpha$) as a function of tetragonal distortion. 
In the cubic limit $\phi\equiv\alpha$, {\it i.e.}, the spins simply rotate 
together with the oxygen octahedra. This suggests 
a strong magnetoelastic coupling in Sr$_2$IrO$_4$, and related phonon 
anomalies at the magnetic transition. The elongation $c>a$ 
(compression $c<a$) of octahedra leads to a decrease (increase) of $\phi$ 
and hence FM moment. At large $c/a$ ratio, $\Gamma_1$ changes sign. 
This marks a spin-flop transition to collinear order along 
the $z$-axis, which happens at $\theta=\pi/4$, {\it i.e.}, 
$\Delta=\lambda/2$. This gives an upper estimate for the tetragonal splitting 
$\Delta<\lambda/2\simeq 190$~meV in Sr$_2$IrO$_4$, which agrees 
with optical data \cite{Moo06,Kim08}. Further, in the cubic limit  we find
$\Gamma_1/{\tilde J}\simeq 0.04$, which is much larger than that in
La$_2$CuO$_4$, and far exceeds possible interlayer interactions \cite{Kei92}. 
This suggests that $XY$-anisotropy is chiefly responsible for finite transition
temperature in Sr$_2$IrO$_4$. From experimental value 
$T_N=240$ K  we estimate ${\tilde J}\simeq45$ meV \cite{note5}, 
which is a realistic value for a $t_{2g}$-system \cite{Kha05}.

We focused above on the Mott insulators \cite{Zaa85}, where the energy 
$\Delta_{pd}$ for a charge transfer from an oxygen to a TM ion 
is larger than $U$. Optical data show that 
the $p$-$d$ transitions in Sr${_2}$IrO$_4$ 
are indeed located at much higher energy than $d$-$d$ ones \cite{Moo06,Kim08}.
We, therefore, neglected the processes with two holes on the oxygen sites \cite{note4}.

To conclude, we have considered magnetic interactions in Mott insulators with 
strong spin-orbit coupling. We find that the symmetry of low energy 
Hamiltonians is dictated by lattice geometry, opening a possibility to 
design exotic spin models like quantum compass and Kitaev models. 
Magnetic properties of the iridate Sr$_2$IrO$_4$ are explained. 
In general, spin-orbit coupled Mott insulators present an interesting 
new class of frustrated systems where the orbital, spin, and 
geometrical frustrations are superimposed via the spin-orbital entanglement, 
giving rise to unusual symmetries of interactions. 

We would like to thank B. Keimer, D.I.~Khomskii, and J. Chaloupka 
for stimulating discussions. G.J. acknowledges support by GNSF 
Grant No. 06-81-4-100.

%%%%%%%%%%%%%%%%%%%%%%%%%%%%%%%%%%%%%%%%%%%%%%%%%%%%%%%%%%%%%%%%%


\begin{thebibliography}{20}
\bibitem{note4} For completeness, we notice that in the case of 
a 180$^\circ$-bond, the $p$-$d$ charge transfer contributions do not change 
the Hamiltonian (\ref{eq2}) and only lead to a trivial 
renormalization of the coupling constants. Conversely, on a 90$^\circ$-bond, 
such processes generate a term 
$J'(\vec S_i\cdot \vec S_{j}-2S_{i}^{\gamma}S_{j}^{\gamma})$, composed 
of Heisenberg and compass models (see also Ref.~\onlinecite{Che08}).   
We find $J'=\frac{1}{9}\frac{U_p}{\Delta_{pd}+U_p/2}\frac{4t^2}{\Delta_{pd}}$ 
which is small in the Mott insulator regime of $\Delta_{pd}\gg U_p,U$.
\end{thebibliography}
\end{document}